\documentclass[twocolumn,amstext,amsmath,amssymb,prl,preprintnumbers,superscriptaddress,floatfix,showpacs,nofootinbib]{revtex4}

\usepackage[dvips]{epsfig}
\usepackage{slashed}
\usepackage{mathrsfs}
\usepackage{graphicx}
\usepackage{xcolor}
\usepackage{hyperref}
\usepackage{dcolumn}
\usepackage{subfigure}
\usepackage{xspace}
\usepackage{ulem}

\begin{document}
\preprint{TUM-HEP-1133/18}
\preprint{KIAS-P18015}

\title{Neutrino Masses from Planck-Scale Lepton Number Breaking}

\author{Alejandro Ibarra}
\affiliation{Physik-Department, Technische Universit\"at M\"unchen, James-Franck-Stra\ss{}e, 85748 Garching, Germany}
\affiliation{School of Physics, Korea Institute for Advanced Study, Seoul 02455, South Korea}

\author{Patrick Strobl}
\affiliation{Physik-Department, Technische Universit\"at M\"unchen, James-Franck-Stra\ss{}e, 85748 Garching, Germany}

\author{Takashi Toma}
\affiliation{Physik-Department, Technische Universit\"at M\"unchen, James-Franck-Stra\ss{}e, 85748 Garching, Germany}


\begin{abstract}
We consider an extension of the Standard Model by right-handed neutrinos and we argue that, under plausible assumptions, a neutrino mass of ${\cal O}(0.1)\,{\rm eV}$ is naturally generated by the breaking of the lepton number at the Planck scale, possibly by gravitational effects, without the necessity of introducing new mass scales in the model.  Some implications of this framework are also briefly discussed.
\end{abstract}

\maketitle

\section{Introduction}
\label{sec:intro}

The masses of the third generation electrically charged fermions are known to a fairly high precision: the top quark mass is $m_t=173.1\pm 0.6$ GeV, the bottom quark mass is $4.18^{+0.04}_{-0.03}$ GeV and the tau lepton mass is $1776.86\pm 0.12$ MeV~\cite{Patrignani:2016xqp}. In the framework of the Standard Model, these masses are generated by postulating an ${\cal O}(1)$ top-Yukawa coupling and ${\cal O}(0.01)$ bottom- and tau-Yukawa couplings to the Higgs field, which lead after the spontaneous electroweak symmetry breaking to the measured masses. The mass of the third generation neutrino, on the other hand, is not positively known. However, oscillation experiments ~\cite{Esteban:2016qun,Capozzi:2017ipn}, measurements of the end point of the tritium $\beta$-decay spectrum~\cite{Kraus:2004zw,Otten:2008zz}, searches for neutrinoless double beta decay~\cite{KamLAND-Zen:2016pfg,Agostini:2013mzu}, and measurements of the cosmic microwave background and cosmological large-scale structure~\cite{Abazajian:2016hbv}, indicate that it should be in the sub-eV range.

The huge hierarchy between the third generation charged fermion and  neutrino masses, at least nine orders of magnitude, suggests the existence of a different mechanism of mass generation for the neutral fermions, other than just a Yukawa coupling to the Higgs field. Arguably, the simplest and most economical framework to explain the differences between the electrically charged fermion masses and the neutrino masses is the seesaw mechanism~\cite{Minkowski:1977sc,Mohapatra:1979ia,Yanagida:1979as,GellMann:1980vs,Schechter:1980gr}. The gauge symmetry of the model allows a (lepton number breaking) Majorana mass term for the right-handed neutrinos, possibly much larger than the electroweak scale, as well as a Yukawa coupling of the lepton doublets and the right-handed neutrinos to the Standard Model Higgs doublet, which generates after electroweak symmetry breaking a (lepton number conserving) Dirac mass term. The interplay between the heavy Majorana mass and the Dirac mass leads to a neutrino mass eigenstate which can be naturally much lighter than the Dirac mass. In this way, the smallness of the neutrino mass can be related to the breaking of the lepton number at very high energies. 

The seesaw model, on the other hand, cannot predict the concrete value of the neutrino masses, as the Yukawa couplings and the right-handed Majorana masses are {\it a priori} undetermined. Furthermore, in contrast to the Standard Model, the parameters of the Lagrangian cannot be determined univocally from experiments, as right-handed neutrinos are not present in the low energy particle spectrum. On the other hand, they are constrained by the requirement of reproducing the measured neutrino oscillation parameters. For instance, assuming a Yukawa coupling of ${\cal O}(1)$, reproducing a neutrino mass scale $\sim 0.1\,{\rm eV}$ requires to postulate a right-handed neutrino mass $\sim 10^{14}\,{\rm GeV}$.  

A large mass scale for the right-handed neutrinos could be generated, {\it e.g.} from a Yukawa interaction with a singlet scalar that takes a large vacuum expectation value~\cite{Chikashige:1980ui,Ballesteros:2016euj}, or from a Planck suppressed dimension-5 operator which couples two right-handed neutrinos to two scalar fields that acquire an expectation value of the order of the grand unification scale $\sim 10^{16}\,{\rm GeV}$~\cite{Barbieri:1979hc}. Alternatively, the $\sim 0.1\,{\rm eV}$ scale could be generated in variants of the seesaw framework, such as in scenarios with warped extra dimensions \cite{Grossman:1999ra,Huber:2003sf,Agashe:2015izu}, or in scenarios with low scale lepton number breaking \cite{Mohapatra:1986aw,Mohapatra:1986bd,Malinsky:2005bi,Kersten:2007vk}.
Whereas these are  phenomenologically acceptable explanations, they are not fully satisfactory from the theoretical point of view, since in order to explain the origin of the $0.1$ eV mass scale it has been necessary to introduce {\it ad hoc} another mass scale in the theory (and possibly new fields).

In the most minimal setup, only two mass scales are available:  the Higgs mass parameter (or alternatively the electroweak symmetry breaking scale) and the  Planck mass, $M_{\rm P}\simeq 1.2\times 10^{19}\,{\rm GeV}$. It has been argued that gravity effects do not preserve global symmetries, such as lepton number~\cite{Abbott:1989jw,Kallosh:1995hi,Hawking:1974sw}. Therefore, a natural value for the right-handed Majorana neutrino mass is around the Planck scale. Under this reasonable assumption, and assuming also an ${\cal O}(1)$ neutrino Yukawa coupling, the predicted neutrino mass is $m_\nu\sim 10^{-6}\,{\rm eV}$~\cite{Barbieri:1979hc}, far too small to explain neutrino oscillation experiments.

In this Letter, we argue that the seesaw model with Planck scale lepton number breaking can naturally generate, under simple and plausible assumptions, an ${\cal O}(0.1)\,{\rm eV}$-neutrino mass scale. To this end,  we extend the Standard Model with two right-handed neutrinos, and we assume that at the cutoff scale of the model (which we identify with the Planck scale), one of the right-handed neutrinos has a mass around the Planck scale, while the other right-handed neutrino is massless. In this way, the only fundamental parameter of the model that breaks lepton number has a Planck scale size. We show that, in general, quantum effects induce a mass for the lighter right-handed neutrino. Furthermore, if the heaviest right-handed neutrino interacts with the left-handed leptons with an ${\cal O}(1)$ Yukawa coupling, then the seesaw mechanism generates an effective neutrino mass with size ${\cal O}(0.1)\,{\rm eV}$, in qualitative agreement with experiments, fairly independently of the value of the Yukawa couplings of the lighter right-handed neutrino. 

\section{Quantum effects on right-handed neutrino masses}
\label{sec:1_gen}

We consider for simplicity a model with one generation of lepton doublets, $L$, and two right-handed neutrinos, $N_1$ and $N_2$. The part of the Lagrangian involving the right-handed neutrinos reads:
\begin{align}
{\cal L}_{N}=\frac{1}{2}\overline{N_i^c}i\slashed{\partial}N_i
- Y_{i} \bar L \widetilde H N_i-\frac{1}{2}M_i \overline{N_i^c} N_i+{\rm h.c.}\;,
\label{eq:Lagrangian}
\end{align}
where $\widetilde H= i \tau_2 H^*$, with $H$ the Standard Model Higgs doublet. We take as cutoff of the theory the Planck scale and we assume that the parameters of the model at that scale are $M_2={\cal O}( M_{\rm P})$, $M_1=0$, such that the lepton number breaking occurs at the Planck scale, possibly by gravity effects, and we assume $Y_2={\cal O}(1)$. We leave $Y_1$ unspecified.

If quantum effects were neglected, the model would predict the existence at low energies of a pseudo-Dirac neutrino pair, with masses $m_\nu\simeq Y_1\langle H^0\rangle \pm\frac{1}{2} Y_2^2 \langle H^0\rangle^2/M_2$, where $\langle H^0\rangle\simeq 174\,{\rm GeV}$ is the Higgs vacuum expectation value. This conclusion is, however, completely altered when properly including quantum effects on the right-handed masses. 

We note that when $Y_1,Y_2=0$ and $M_1,M_2=0$  the Lagrangian Eq.~(\ref{eq:Lagrangian}) is invariant under the global $U(1)_L$ transformation $L\rightarrow e^{i\alpha_L} L$ and $U(2)_N$ transformation $N\rightarrow V N$, with $N=(N_1,N_2)$ and $V$ a $2\times 2$ unitary matrix. However, when setting $Y_1,Y_2\neq 0$ and $M_2\neq 0$, the global $U(2)_N$ symmetry is completely broken, even if $M_1=0$. Therefore, and since there is no symmetry protecting the lightest right-handed neutrino mass against radiative effects, it will be generated via loops, and will be proportional to the order parameter of the lepton number breaking, $M_2$.  We also note that if any of the parameters $Y_1$, $Y_2$, or $M_2$ is equal to zero, the symmetry of the Lagrangian is enhanced and this symmetry will protect $M_1$  against quantum effects. 

\begin{figure}[t]
	\begin{center}
		\includegraphics[width=0.4\textwidth]{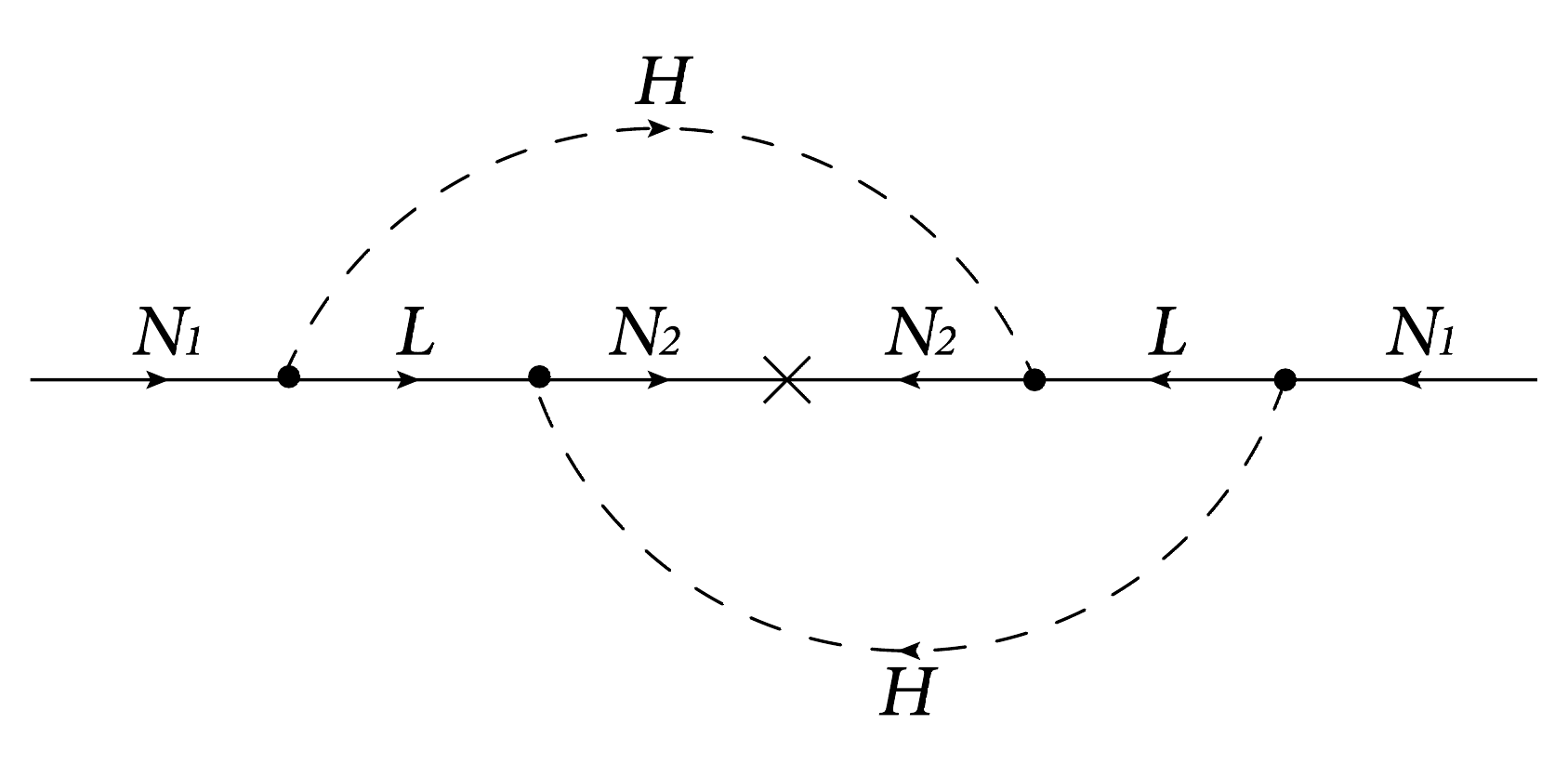} 
	\end{center}
	\caption{\small Two-loop diagram leading to the radiative generation of $M_1$.}
	\label{fig:diagram} 
\end{figure}

An explicit calculation confirms this expectation. At the one-loop level, one finds corrections to the right-handed masses which are proportional to themselves~\cite{Casas:1999tp,Casas:1999tg}, such that $M_1$ remains massless. However, at the two-loop level one finds nonvanishing contributions to $M_1$, through the diagram depicted in Fig.~\ref{fig:diagram}. Concretely, for the toy Lagrangian Eq.~(\ref{eq:Lagrangian}), we find that the right-handed neutrino masses, evaluated at the scale $\mu=M_2$, approximately read:
\begin{align}
M_2|_{\mu=M_2}&\simeq M_2 \;,\nonumber \\
M_1|_{\mu=M_2}&\simeq 
\frac{4 \,Y_1^2 Y_2^2}{(16\pi^2)^2} M_2 \log\left(\frac{M_2}{M_{\rm P}}\right) \;,
\label{eq:M1_at_M2}
\end{align}
where all parameters in the right-handed side of these equations are evaluated at the cutoff scale. As anticipated, the lightest right-handed neutrino mass is proportional to $Y_1$, $Y_2$ and $M_2$, such that it vanishes when any of these is equal to zero. Finally, below the scale $\mu=M_2$, quantum corrections induced by the Yukawa coupling $Y_1$ will modify the value of the lightest right-handed neutrino mass at the scale $\mu=M_1$. These corrections are, however, typically small and will not affect our main conclusions.  

At low energies the heavy neutrinos can be integrated out, leading to an effective neutrino mass
\begin{align}
m_\nu&\simeq\left(\frac{Y_1^2}{M_1}\Big|_{\mu=M_1}+\frac{Y_2^2}{M_2}\Big|_{\mu=M_2}\right) \langle H^0\rangle^2 \nonumber \\
&\simeq\left(\frac{(16\pi^2)^2}{4 Y_2^2 \log(M_2/M_{\rm P})}+Y_2^2\right) \frac{ \langle H^0\rangle^2}{M_2} \;.
\label{eq:nu_mass_1gen}
\end{align}
Here, we have again neglected the effect of quantum corrections between the scale $\mu=M_1$ and the scale of oscillation experiments, first discussed in Refs.~\cite{Babu:1993qv,Chankowski:1993tx}, since they will not affect our conclusions. 

For perturbative values of $Y_2$, namely $Y_2\leq \sqrt{4\pi}$,  the first term in Eq.~(\ref{eq:nu_mass_1gen}) dominates. So, the neutrino mass is mostly generated by the interaction of the lepton doublet with $N_1$ and takes the value
\begin{align}
m_\nu
&
\simeq\left(\frac{(16\pi^2)^2}{4  \log(M_2/M_{\rm P})}\right) \frac{\langle H^0\rangle^2}{Y_2^2 M_2} \nonumber \\
&  \simeq 0.05\, {\rm eV}\,\, \left(\frac{Y_2}{0.6}\right)^{-2} \left(\frac{M_2}{1.2\times 10^{19}\,{\rm GeV}}\right)^{-1}\;,
\label{eq:nu-mass}
\end{align}
which is naturally of the correct size for $M_2\sim M_{\rm P}$ and  $Y_2\sim 0.6$ [here, we have approximated  $\log(M_2/M_{\rm P})\approx 1$]. It is notable that this result holds independently of the value of the Yukawa coupling $Y_1$ (as long as it is nonzero), and correspondingly of the value of the lightest right-handed neutrino mass, which can be either $M_1\sim 10^{14}\,{\rm GeV}$ when $Y_1\sim 1$, as can be checked from Eq.~(\ref{eq:M1_at_M2}), or much lighter, depending on the value of $Y_1$.

\section{The three generation case}

This discussion can be extended to the realistic case of three generations of lepton doublets. Here we just sketch the basic ideas and we defer a detailed discussion to a forthcoming publication~\cite{inprep}.

The neutrino Yukawa coupling, $Y$, and the right-handed neutrino mass
matrix, $M$, are in this case $3\times 3$ complex matrices with
eigenvalues $y_1\leq y_2\leq y_3$, and $M_1\leq M_2\leq M_3$.  Similarly to the two-generation case, the global $U(3)_N$ flavor group can be completely broken even when some of the eigenvalues of the right-handed mass matrix and/or the Yukawa couplings vanish at the cutoff scale. More concretely, it can be checked that, generically, a rank-1 right-handed mass matrix and a rank-2 neutrino Yukawa coupling
already break completely the $U(3)_N$ flavor group. Therefore, quantum corrections will increase the rank of the right-handed neutrino mass matrix from 1 to 3. 
	
We present here the results for the scenario where at the cutoff scale $M_3={\cal O}(M_P)$, $M_1\ll M_2\ll M_3$ (the analysis for the case where  $M_1=M_2=0$, such that two mass eigenvalues are generated by quantum effects, is technically more complicated and will be discussed at length in Ref.~\cite{inprep}). After including quantum effects, we find a right-handed neutrino mass spectrum  which approximately reads  at the scale $\mu=M_3$: 
\begin{align}
M_3|_{\mu=M_3}& \sim M_3\;, \nonumber \\
M_2|_{\mu=M_3}&\sim -\frac{1}{(16\pi^2)^2} M_3 y_3^4 \sin^4\zeta\times {\cal O}(1){\rm~ factors} \;, \nonumber \\
M_1|_{\mu=M_3} &\sim M_2 \frac{\sin^4 \xi}{ \sin^4\zeta}\times {\cal O}(1){\rm~ factors} \;,
\label{eq:RH-spectrum-rank2}
\end{align}
where $\zeta$ and $\xi$ are combinations of mixing angles in the right-handed neutrino sector. We note that the result is independent of $M_1$ (which is washed-out by quantum effects), and that the mass scale of two eigenstates is determined by $M_3={\cal O}(M_P)$. Below the scale $\mu=M_3$ the two lighter right-handed masses are subject to further quantum effects, but their values at the corresponding decoupling scales $M_2|_{\mu=M_2}$ and  $M_1|_{\mu=M_1}$ do not differ significantly from those in Eq.~(\ref{eq:RH-spectrum-rank2}). 

The active neutrino masses are generated by the seesaw mechanism. Similarly to the two-generation case, one obtains one neutrino mass eigenvalue with size
\begin{align}
m_{\nu_a}&\sim\frac{y_3^2}{M_2}\Big|_{\mu=M_2} \langle H^0\rangle^2  
\nonumber\\
& \sim {\cal O}(0.1)\,{\rm eV}\,\, \left(\frac{y_3}{1}\right)^{-2} \left(\frac{M_3}{1.2\times 10^{19}\,{\rm GeV}}\right)^{-1}\;,
\end{align}
with a very mild sensitivity to the values of $M_1, M_2, y_1$ and $y_2$ at the cutoff scale, but with some sensitivity to the values of the right-handed mixing angles; in this estimate, we have assumed that all the relevant mixing angles are sizable. In this scenario, therefore, one of the neutrino masses is generically predicted to be ${\cal O}(0.1)$ eV, in qualitative agreement with experiments.

In contrast, the values of the two lighter active neutrino masses are not predicted, and strongly depend on the free parameters $M_1, M_2, y_1$ and $y_2$, as well as on the right-handed mixing angles. For generic mixing angles, one finds  
\begin{align}
m_{\nu_b}\sim \frac{y_2^2}{M_1}\Big|_{\mu=M_1} \langle H^0\rangle^2  \;,~
m_{\nu_1}\sim \frac{y_1^2}{M_3}\Big|_{\mu=M_3} \langle H^0\rangle^2 \;.
\end{align}
Clearly $m_{\nu_1}\ll 10^{-6}$ eV and, correspondingly, the framework predicts a hierarchical neutrino mass spectrum, with $\nu_1$ the lightest
eigenstate. On the other hand, $m_{\nu_b}$ can be larger or smaller than
$m_{\nu_a}$, depending on the parameters, {\it i.e.} $a=3$ and $b=2$, or
vice versa, with the standard labeling of the neutrino mass
eigenstates. In either case, it is always possible to find high energy
parameters such that the observed hierarchy between the mass splittings $|\Delta m_{31}^2|/\Delta m_{21}^2\simeq 30$ is
reproduced. 

\begin{figure}[t]
	\begin{center}
	 \includegraphics[width=0.47\textwidth]{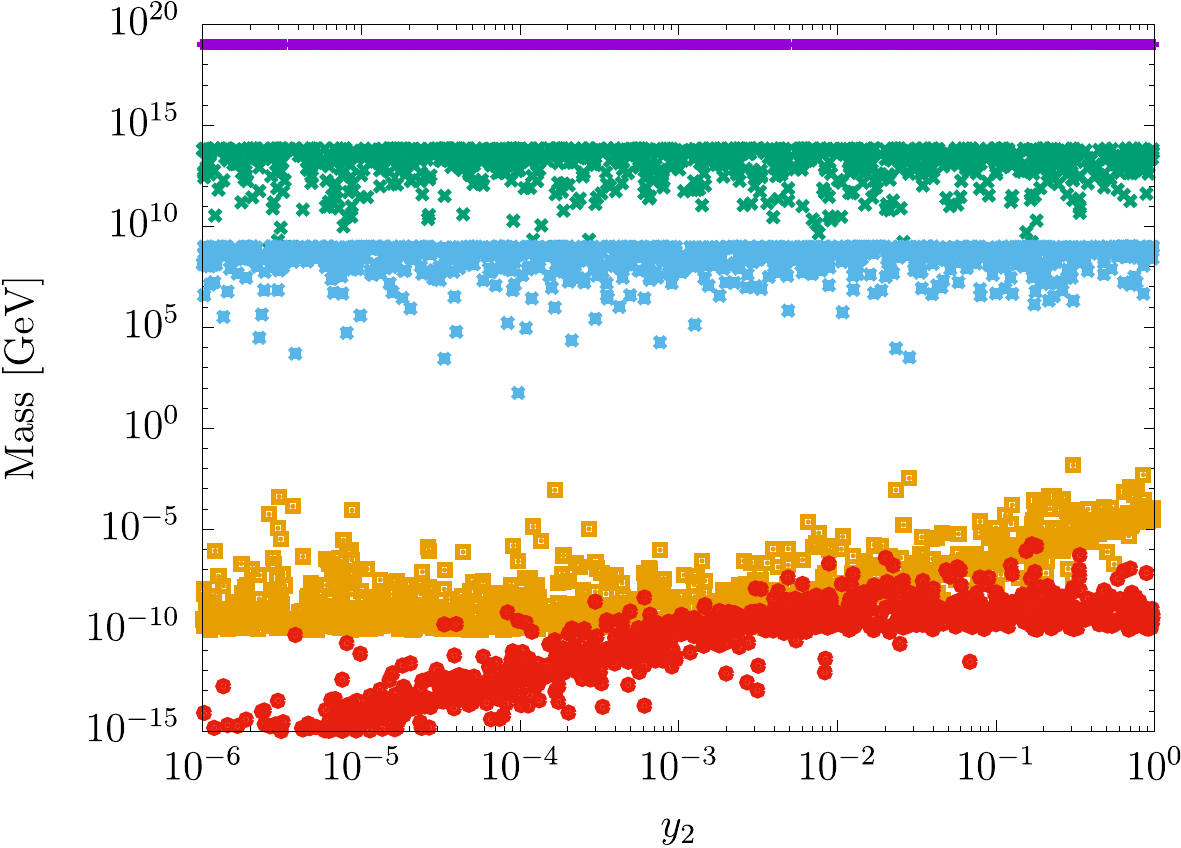} 
	\end{center}
	\caption{\small Physical neutrino masses as a function of
 $y_2$ for the scenario where at the cutoff scale $M_3=10^{19}$ GeV,
 $M_2=10^9$ GeV, $M_1=0$, $y_3=1$, $y_1=0$, and the right-handed mixing angles take random values between 0 and $2\pi$. The lightest neutrino mass is $\ll 10^{-6}$ eV and lies outside of the figure.}
	\label{fig:masses} 
\end{figure}

This is illustrated in  Fig.~\ref{fig:masses}, which shows
the neutrino mass eigenstates as a function of the next-to-largest
neutrino eigenvalue $y_2$, for the specific case $M_3=10^{19}$ GeV,
$M_2=10^9$ GeV, $M_1=0$, $y_3=1$, $y_1=0$ at the cutoff, taking right-handed mixing angles with random values between 0 and $2\pi$. We find that for arbitrary choices of parameters the framework tends to produce a too large hierarchy between the atmospheric and the solar neutrino mass scales; this is generic of the seesaw mechanism~\cite{Casas:2006hf} and not specific to our framework. Yet, we find that it is possible to reproduce the observed mass hierarchy for particular choices (not necessarily fine-tuned) of the right-handed mixing angles and $y_2$, in this case provided  $y_2\sim 10^{-4}-10^{-1}$. More importantly, our framework makes the nontrivial prediction of a neutrino mass eigenvalue in the ballpark of neutrino experiments, regardless of the values of $M_1,M_2, y_1,y_2$ and with fairly mild assumptions about $M_3$ and $y_3$. This is the main feature of our framework and the main result of this Letter.  We note that a mild neutrino mass hierarchy could be naturally reproduced by extending the scalar sector of the model, using the mechanism discussed in Refs.~\cite{Ibarra:2011gn,Grimus:1999wm}). A detailed analysis will be presented elsewhere.

Our assumption for a strongly hierarchical right-handed neutrino at the Planck scale is purely phenomenological. Nonetheless, it is interesting to speculate about the possible origin of such structure. If the right-handed neutrino mass matrix is of gravitational origin, so that one can relate the lepton number breaking to physics at the Planck scale, it is natural to identify the overall right-handed neutrino mass scale with the Planck scale. Also, it has been argued that a gravitationally induced effective neutrino mass matrix would take a ``democratic" form \cite{Akhmedov:1992hh}. The same rationale applied to a gravitationally induced right-handed neutrino mass would yield
\begin{align}
M(\Lambda) = \omega M_{\rm P} \begin{pmatrix} 1 & 1 & 1 \\ 1 & 1 & 1 \\ 1 & 1 & 1 
\end{pmatrix}\;,
\label{eq:grav-induced-RH-mass}
\end{align}
with $\omega={\cal O}(1)$, which has eigenvalues $\omega M_{\rm
P}\{0,0,3\}$. Furthermore, it was argued in Ref.~ \cite{Berezinsky:2004zb} that the exact ``democratic" form might be perturbed  by topological
fluctuations at the Planck scale~\cite{Coleman:1988cy,Giddings:1988cx}. In this case, the right-handed neutrino mass spectrum at the cutoff would consist of a Planck-scale
eigenvalue, and two nonzero eigenvalues, plausibly much lighter than the Planck scale, as assumed in our work. On the other hand, the fundamental mechanism generating Yukawa couplings remains a mystery to this day. However, in view of the observed quark and charged lepton Yukawa couplings, it is feasible that the neutrino Yukawa matrix has rank larger than 1, that the largest eigenvalue is sizable, and that mixes different generations. 

\section{Outlook}

In this framework, one (or two) of the right-handed masses are determined purely from quantum effects. Therefore, this framework renders a higher predictivity. More specifically, for the toy model with one lepton doublet and two right-handed neutrinos, the most general framework is defined by 5 free parameters at the cutoff scale: 2 moduli and 1 phase in the Yukawa coupling, and 2 right-handed masses. However, under the assumption that one of the right-handed masses vanishes at the cutoff scale (or is negligible compared to the radiatively induced contribution to the mass), all observables will depend, at most, on 3 free parameters, as one phase can be rotated away by a field redefinition. Notably, some observables may depend on even less parameters, such as the active neutrino mass, which only depends on two free parameters [{\it cf.} Eq.~(\ref{eq:nu-mass})]. For the  realistic case with a rank-3 Yukawa matrix and a rank-3 right-handed neutrino mass matrix, the most general framework is defined by  18 free parameters: 9 moduli and 6 phases in the Yukawa matrix, as well as the 3 right-handed masses. Also in this case, lowering the rank of the right-handed neutrino mass matrix leads to a reduction of the number of parameters relevant to calculate observable quantities. Correspondingly, the predictivity of the model is enhanced, especially in regard to the number of unknown phases, which could have implications for testing leptogenesis with low-energy observables~\cite{inprep}.  

In this work we have been motivated by generating dynamically the  right-handed neutrino mass scale necessary to reproduce the neutrino mass inferred from oscillation experiments. However, a similar rationale can be applied in other frameworks to generate a mass scale for a fermion singlet. For instance, several works have advocated a keV mass sterile neutrino as dark matter candidate~\cite{Dodelson:1993je,Shi:1998km}. This choice was purely based on various phenomenological considerations, but lacked theoretical justification.  Using the mechanism presented in this work, the keV mass scale could be related to the breaking of lepton number at very high energies, which is transmitted via loops to an initially massless right-handed neutrino through a small Yukawa coupling [namely, $Y_1$ in Eq.~(\ref{eq:M1_at_M2})]. In this way, no new mass scales have to be introduced in the model. 

Finally, we would like to emphasize that quantum effects on the right-handed neutrino masses can also be important in frameworks where the scale of lepton number breaking is below the Planck scale (as in some grand unification models) and/or in any framework where the tree-level right-handed mass hierarchy is very large. In this case, quantum effects induced by the heavier mass eigenvalues can radiatively induce masses for the lighter eigenvalues which can be much larger than the tree-level values, possibly affecting the phenomenology of the model. 

\section{Conclusions}
We have considered an extension of the Standard Model by a Planck scale mass right-handed neutrino, motivated by the fact that the lepton symmetry is likely to be broken by gravitational effects at the Planck scale. 
We have argued that, in general, the masses of the lighter right-handed neutrinos are not protected by any symmetry and therefore they receive sizable contributions, possibly dominant, from quantum effects induced by the Planck-scale mass right-handed neutrino. Concretely, we have explicitly shown that an initially massless right-handed neutrino becomes massive due to two-loop effects. Furthermore, we have shown that when the heaviest right-handed neutrino interacts with the lepton doublets with an ${\cal O}(1)$ Yukawa coupling, the seesaw mechanism generates an effective neutrino mass which is naturally of ${\cal O}(0.1)\,{\rm eV}$, as suggested by neutrino oscillation experiments. This result supports the seesaw mechanism with Planck scale lepton number breaking as the origin of the observed neutrino mass scale.

\vspace{0.5cm}
{\it Acknowledgements}: 
This work has been partially supported by the DFG cluster of excellence EXC 153 ``Origin and Structure of the Universe'' and by the Collaborative Research Center SFB1258. T.T. acknowledges support from JSPS Fellowships for Research Abroad.


\bibliography{references}

\end{document}